\documentclass{INTERSPEECH2023}

\usepackage{makecell}

\interspeechcameraready 

\title{Investigating Range-Equalizing Bias in Mean Opinion Score Ratings of Synthesized Speech}
\name{Erica Cooper, Junichi Yamagishi}
\address{
  National Institute of Informatics, Japan}
\email{ecooper@nii.ac.jp, jyamagis@nii.ac.jp}

\begin{document}

\maketitle
 
\begin{abstract}
Mean Opinion Score (MOS) is a popular measure for evaluating synthesized speech.  However, the scores obtained in MOS tests are heavily dependent upon many contextual factors.  One such factor is the overall range of quality of the samples presented in the test -- listeners tend to try to use the entire range of scoring options available to them regardless of this, a phenomenon which is known as {\em range-equalizing bias.}  In this paper, we systematically investigate the effects of range-equalizing bias on MOS tests for synthesized speech by conducting a series of listening tests in which we progressively ``zoom in'' on a smaller number of systems in the higher-quality range.  This allows us to better understand and quantify the effects of range-equalizing bias in MOS tests.

\end{abstract}

\section{Introduction}

Listening tests where human listeners rate the quality of generated samples are the gold standard for evaluating models of synthesized speech such as text-to-speech or voice conversion, and the Mean Opinion Score (MOS) test \cite{ITUMOS}, where listeners rate audio samples one by one on an ordinal scale, is a popular testing paradigm for rating and comparing different approaches.  However, MOS tests are known to be very context-dependent, where the context includes the demographics of the listeners that were recruited to participate in the test, their personal opinions and individual differences, the environment in which they are listening to samples, the types of synthesis systems included in the test, the sentences chosen to synthesize test material, and even the appearance of the testing interface.  In this paper we investigate the effect of {\em range equalizing bias,} which is the tendency of listeners to use the entire range of choices on the rating scale, regardless of the absolute quality of the samples present in the test.  To do this, we start with an existing large-scale database of audio samples and their MOS ratings, and conduct new listening tests where we progressively ``zoom in'' on the high-quality region of synthesis systems.  By quantifying the ways in which MOS ratings change as listening test context is changed systematically, we can better understand the effects of context on listening test results, which can help researchers to design their listening tests in more principled ways, and which can also help to improve automatic MOS prediction.

The BVCC dataset \cite{BVCC} is a large-scale dataset of MOS ratings for samples from 187 different text-to-speech and voice conversion systems covering over a decade of progress in these fields.  These samples were evaluated all together in one large-scale MOS test.  This dataset provided for the first time a huge variety of synthesized samples all evaluated together in the same context, and has since been used as training material for automatic MOS predictors \cite{generalization-mos,huang22voicemos}.  However, this dataset has several shortcomings.  The wide variety of synthesis methods present in the dataset provide good coverage and generalizability \cite{generalization-mos}, but this also means that there are fewer fine-grained and significant differences to be found between synthesizers of similar quality, and that the ability of MOS predictors trained on this data to correctly rank synthesis systems that are close in quality will be impaired.  
The number of systems is also much larger than a typical or representative listening test. Furthermore, present-day researchers are less likely to want to compare their proposed synthesis methods with ones from many years ago. Present-day, high-quality synthesis systems are the more relevant comparisons, and there is especially room for improvement in this regard as shown from the results of the VoiceMOS Challenge -- while low-quality samples were consistently easy to predict for all teams, middle-range and high-quality systems remained more difficult to predict accurately.  These points motivated us to revisit the BVCC dataset and conduct new listening tests on progressively smaller subsets of the synthesis systems, focusing in towards the top of the range of quality.  Data from these new listening tests allow us to observe and quantify the effects of range-equalizing bias on MOS of synthesized speech.  We expect to see scores for some systems to decrease as listeners attempt to use the entire range of score choices, and we also expect to be able to find more significant differences between systems.  

The distributions of scores reveal the presence of range-equalizing bias in MOS listening tests for synthesized speech, while the skew of the distributions demonstrate that listeners nevertheless do maintain some sense of absolute quality while completing the test.  We can see more significant differences emerge as the range of synthesis systems becomes smaller, and we can also observe a substantial decrease of about one MOS point in scores for some systems.  In Section \ref{sec:relatedwork} we discuss some other studies which address context-dependency and bias in perceptual tests.  In Section \ref{sec:data} we describe our data and listening test design.  Section \ref{sec:results} describes the results of our listening tests and analysis of what happens when a smaller range of synthesis systems is evaluted.  We conclude in Section \ref{sec:conclusions}.

\section{Related work}
\label{sec:relatedwork}

A comprehensive review of biases in listening tests \cite{zielinski2008some} documents the various factors that can affect listener ratings.  These include recency, listener expectations, stimulus frequency, perceptually non-equal spacing between the words chosen for the rating categories that may vary by language, and even factors such as the appearance of the testing equipment or user interface.  The range-equalizing bias effect is described as a ``rubber ruler,'' meaning that listeners tend to use the entire available range of choices, regardless of ``absolute'' quality.  The authors recommend using anchoring techniques to counteract this.

However, even listening test paradigms that use anchoring are not completely immune to this type of bias.  The same authors \cite{zielinski2007potential} systematically varied the samples in a MUSHRA (Multiple Stimuli with Hidden Reference and Anchor) test \cite{mushra}, in which listeners are presented with samples from several systems including a reference sample and one or more anchor samples, finding that even when the anchors are held constant, the scores can change depending on whether there are more higher-quality or lower-quality samples present in the test.

The presentation and context of  the audio sample being evaluated can affect the scores, even when the sample itself stays the same -- scores can differ whether a sample under consideration is presented in isolation, or in a paragraph or dialogue context, and these scores do not even necessarily correlate with each other \cite{clarkevaluating}.   Scores can also be affected by the wording of the question or instructions being given to the listeners \cite{dall2014rating,o2021factors}.  

Listeners may also differ in their personal preferences, and it is therefore important to collect enough different opinions.  In a revisiting of the 2013 Blizzard Challenge \cite{blizzard2013} listening test data, the rank order and number of significant differences were found to change depending on sentence coverage and number of listeners, and statistical significance only stabilized at 30 listeners \cite{wester2015we}.  Another revisiting of Blizzard 2013 re-evaluated three representative systems from that challenge alongside four modern neural ones, finding that MOS ratings for the three older systems dropped a full point when evaluated together with the modern systems, even though the rankings were preserved \cite{le2022back}.  Furthermore, a significant difference between two of the older systems was lost in the new test, indicating that their differences became ``compressed'' by the presence of the better systems.  

These prior works emphasize that MOS is a relative rather than absolute measure, and that ratings heavily depend on many factors including the overall range of quality of the samples presented in the test.  In this work, we aim to more systematically investigate and quantify how MOS changes as the overall quality of samples in the test changes, and we also aim to better understand and characterize the differences between listening tests that have a large vs. small range of qualities represented.

\section{Data and listening test design}
\label{sec:data}

\subsection{The BVCC dataset}

We used audio samples from the BVCC dataset \cite{BVCC}, a large-scale set of MOS ratings for samples from 187 different systems (including reference natural speech) from past Blizzard Challenges \cite{king2014measuring}, Voice Conversion Challenges \cite{vcc2020}, and ESPnet-TTS \cite{hayashi2019espnettts}.  These MOS ratings have been made public by the organizers of the VoiceMOS Challenge \cite{huang22voicemos}.  The corresponding audio samples are all sampled at 16 kHz, and have been amplitude-normalized using sv56 \cite{sv56}.  38 samples per synthesis system were chosen for BVCC, balancing for genre (e.g., news, audiobook, conversational) and source/target speaker (for voice conversion systems) where relevant.  

\subsection{New listening tests}

We conducted new listening tests to re-evaluate these synthesis systems in different contexts, ``zooming in'' on subsequently smaller, higher-quality regions of the original BVCC listening test data.  We sorted the 187 systems by their MOS ratings in the original BVCC data, and created four subsets by approximately halving the number of systems each time, keeping the highest-rated half according to their original ratings.  32 samples per system were selected for all tests out of the original 38 samples, maintaining genre/speaker balance per challenge.  The listening test sets (lists of samples to be rated by one listener) were designed to contain about the same number of samples regardless of the zoom level, which we accomplished by doubling the number of samples from a given system that appear in one set every time the number of systems is halved.  The entire test was designed to be taken by 200 unique listeners who are native speakers of English, defined as having spoken English as their first language since birth.  Our test design has fewer ratings per sample for the more zoomed-out tests and smaller numbers of listeners for the more zoomed-in tests as a necessary tradeoff due to budgetary constraints.  In a pre-test questionnaire, we asked listeners for their age range, gender, whether or not they have any hearing impairment, and their dialect of English.  The question asked to listeners for each sample was, {\em ``The voice you hear is a computer-generated voice (synthesized voice). While some synthetic voices sound high quality and natural, some of them may be slightly degraded by computer processing and sound artificial, or their voices and inflections may sound discontinuous or unnatural. Please rate the overall quality of the synthesized voice subjectively on a 5-point scale from `very good' to `very bad'.''}  The rating options were {\em Very Bad, Bad, Fair, Good,} and {\em Very Good}, corresponding to numerical MOS ratings from 1-5. Ethics approval was obtained from our institution.  Listeners were not permitted to complete more than one set from any of the zoom levels, to obtain ratings from a larger variety of listeners and to avoid cross-contamination of the different listening test contexts of each zoom level.    
Details of the four zoomed-in test parts, as well as the original BVCC listening test for comparison, are in Table \ref{tab:LTinfo}.  The 6\% zoom test contains the top 11 systems from the original BVCC, and we note that five of these are in fact natural speech.

\setlength{\tabcolsep}{5.5pt}
\begin{table}[th]
  \caption{Details of listening test design.  Information for the original BVCC listening test from \cite{BVCC,generalization-mos} is included as the 100\% zoom level for comparison.}
  \label{tab:LTinfo}
  \centering
  \begin{tabular}{cccccc}
    \toprule
   \thead{Zoom \\ level} & \thead{Number \\ of \\ systems} & \thead{Samples \\ per system \\ in a set} & \thead{Number \\ of \\ listeners} & \thead{Ratings \\ per \\ sample} & \thead{Ratings \\ per \\ system}\\
    \midrule
    \em{100\%}  & \em{187} & \em{1} & \em{304} & \em{8} & \em{304} \\
    50\%   & 93  & 2 & 64 & 4 & 128 \\
    25\%   & 46  & 4 & 64 & 8 & 256 \\
    12\%   & 23  & 8 & 48 & 12 & 384 \\
    6\%    & 11  & 16 & 24 & 12 & 384 \\
    \bottomrule
  \end{tabular}
  
\end{table}

\section{Results}

\label{sec:results}

We can observe the change in the distributions of the ratings in the original test compared to the new tests, the emergence of more statistically-significant differences, and an overall trend of scores decreasing as we zoom in to the higher-quality region.

\subsection{Distributions of scores of different listening tests}

\begin{figure*}[t]
  \centering
  \includegraphics[width=\linewidth]{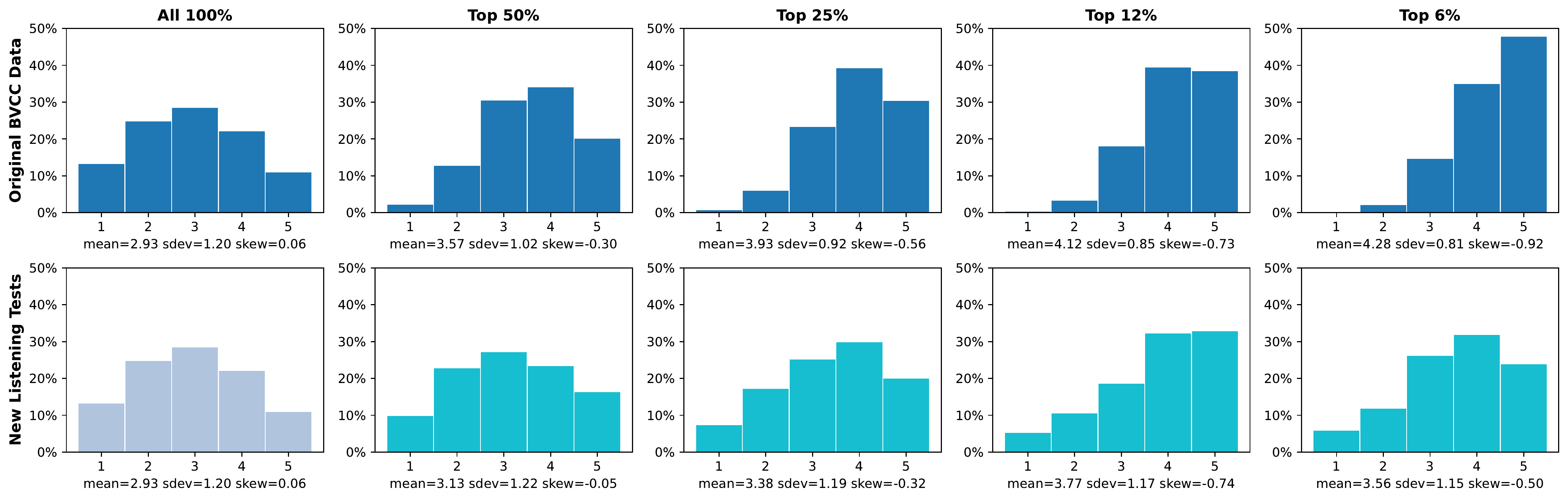}
  \caption{Distributions of listener scores for subsets of the original BVCC dataset (top), and new zoomed-in listening tests (bottom).  The 100\% zoom test was not repeated in our new listening tests, but we copy it in the bottom row for easier visual comparison.  Bar heights show the percent of ratings with the given label.}
  \label{fig:distrs}
\end{figure*}

Figure \ref{fig:distrs} shows the distributions of the listener ratings for each zoomed-in new listening test (bottom row), and the ratings for the same subset of systems in the original BVCC dataset (top row) where all listeners heard samples from all 187 systems.  Statistics for mean, standard deviation, and skew for each of these distributions are also shown.  We can observe the presence of range-equalizing bias at the higher zoom levels -- while no samples from the top 11 systems (6\% zoom) received ratings of 1 in the original BVCC listening test, we nevertheless observe many ``1'' ratings in the new test where listeners {\em only} heard samples from the top 6\% of systems and thus gave the worst ones a rating of 1.  However, we can also observe that listeners nevertheless maintain some sense of absolute quality at the same time that they adjust their ratings to fill the whole scale -- this is demonstrated by the strong skew in the distributions of the new listening tests as we zoom in towards a smaller number of higher-quality samples.  Despite the fact that these listeners have not heard the samples from the lower quality systems, they still select higher scores more frequently.

\subsection{Emergence of significant differences}

As we zoom in towards the top of the range of quality, we expect more significant differences between systems to emerge.  Table \ref{tab:statsig} shows the comparison of the number of statistically-significant differences in the relevant subsets of the original BVCC data, along with the numbers for each new test.  Statistical significances were computed using the Mann-Whitney U test following \cite{Rosenberg2017} with Bonferroni correction for multiple comparisons.  We observe that as the range of qualities of systems under evaluation becomes narrower, the number of statistical differences tends to increase as expected.  This indicates that some differences between systems are being ``compressed'' by including a wider range of systems in the listening test.  We can also observe that the number of statistically-significant pairs of systems in the top 6\% of systems is 25 in both the 12\% listening test and the 6\% test, indicating that all of the perceivable differences may have been found -- this could be further corroborated by doing a pairwise comparison between the top 6\% of systems, as pairwise comparisons can be seen as an extreme case of zooming in.  We leave this for future work.

\setlength{\tabcolsep}{6pt}
\begin{table}[th]
  \caption{Number of statistically-significant pairs of systems that were found in subsets of the original BVCC data and new listening tests.  The rows are listening tests, and the columns say which top percent subset out of the 187 systems are being considered.  The largest number of significant differences for a given percent subset are indicated in \textbf{bold}.}
  \label{tab:statsig}
  \centering
  \begin{tabular}{llllll}
  \toprule
   & 100\%  & 50\% & 25\% & 12\% & 6\% \\
     \midrule
    {\em BVCC}   & \textbf{13068}  & {\em \textbf{2400}} & {\em 354} & {\em 61} & {\em 2} \\   
    50\% test   & -  & 1823 & 381 & 75 & 11 \\ 
    25\% test & - & - & \textbf{562} & 117 & 17 \\
    12\% test & - & - & - & \textbf{152} & \textbf{25} \\
    6\% test  & - & - & - & - & \textbf{25} \\
    \bottomrule
  \end{tabular}
  
\end{table}
\setlength{\tabcolsep}{5.5pt}

\subsection{Correlations}
\label{sec:correl}

There were only two statistically-significant differences among the top 11 systems in the original BVCC listening test data, which raises the possibility that the rankings of systems may change in each test.  We can measure this by looking at correlations, including rank-based correlation measures.  We focus on just the top 11 systems which were present in all listening tests for conciseness, and measure correlations of their scores in each new listening test with their scores in the original BVCC listening test.  We report the following system-level and utterance-level correlation coefficients in Table \ref{tab:correl}: the linear correlation coefficient (LCC), which is basic correlation; Spearman's rank correlation coefficient (SRCC), which considers how much the rankings stayed the same; and Kendall's Tau (KTAU), another ranking-based correlation measure that is more robust to errors.

\begin{figure*}[ht]
  \centering
  \includegraphics[width=\linewidth]{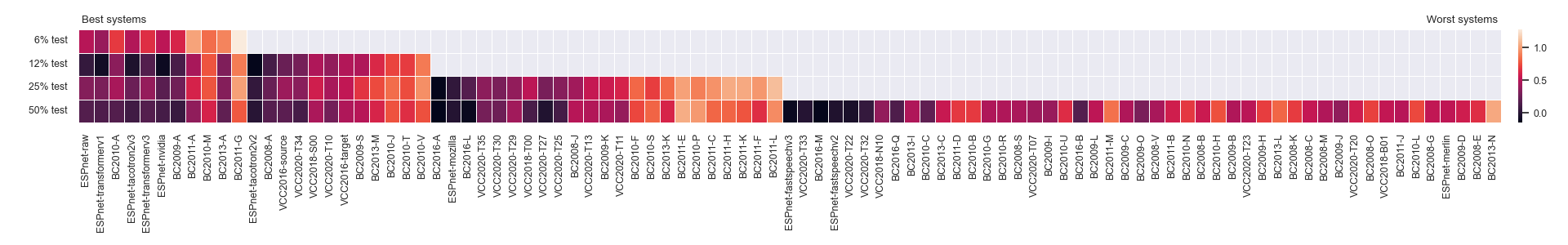}
  \caption{Heatmap of each system's original score in the BVCC listening test, minus its score in the new tests.}
  \label{fig:heatmap}
\end{figure*}

\begin{table}[th]
  \caption{Correlations of scores of top 11 systems in each new listening test compared to original BVCC data.}
  \label{tab:correl}
  \centering
  \begin{tabular}{lllllll}
  \toprule
   & \thead{LCC \\ (sys)}  & \thead{LCC \\ (utt)} & \thead{SRCC \\ (sys)} & \thead{SRCC \\ (utt)} & \thead{KTAU \\ (sys)} & \thead{KTAU \\ (utt)} \\
     \midrule
    50\% test  & .85  & .53 & .83 & .52 & .66 & .38 \\ 
    25\% test & .71 & .38 & .68 & .36 & .48 & .26 \\
    12\% test & .67 & .31 & .63 & .29 & .41 & .20 \\
    6\% test & .42 & .20 & .31 & .16 & .24 & .11 \\
    \bottomrule
  \end{tabular}
  
\end{table}

Correlations decrease substantially as we zoom in.  The two pairs of systems with significant differences in the original BVCC data do retain their order as well as their significant differences, and the systems that switch ranks are ones that did not have significant differences in the original test.

\subsection{Decrease in MOS ratings for some systems}

In a listening test that evaluates a small number of high-quality systems, it is expected that the worst system will have a low score, even if it was rated highly in a test with a wider range of systems -- that is to say, we expect some systems' scores to decrease as we zoom in.  We examine this in Figure \ref{fig:heatmap}, which is a heatmap showing the drop in scores of each system in each new test compared to the original BVCC data.  Darker colors indicate that the score did not change much from the original listening test, and lighter colors indicate a larger score drop.  Gray boxes indicate that a system was not present in a given test.  Synthesis systems are sorted first according to their rank order in the 6\% zoom test, then remaining systems in the 12\% test are sorted according to their order in that test, and so on.  It can be seen that systems that ranked low in their respective tests had larger score drops.  The largest drop was of 1.28 MOS points for system BC2011-G, which was the worst-ranked system in the 6\% zoom-in test.  Several systems had somewhat higher scores in some of the new tests compared to BVCC; these are typically the higher-quality systems such as ESPnet.  The system that had the largest increase in score (0.15 MOS points) was BC2016-A, the professional British English audiobook speech used as training data in the Blizzard Challenge 2016.

In Table \ref{tab:drop}, we report the largest drop in score for each new listening test compared to the original score in BVCC, along with the name of the synthesis system with this score drop.

\setlength{\tabcolsep}{6pt}
\begin{table}[th]
  \caption{Synthesis systems with the largest drops in MOS in each new listening test compared to the original BVCC.}
  \label{tab:drop}
  \centering
  \begin{tabular}{lll}
  \toprule
   \thead{Test}  & \thead{Largest \\ score drop} & \thead{System ID} \\
   \midrule
    50\% test  & 1.05 & BC2011-E \\ 
    25\% test & 1.11  & BC2011-L \\
    12\% test & 0.89 & BC2011-G \\
    6\% test & 1.28 & BC2011-G \\
    \bottomrule
  \end{tabular}
  
\end{table}

Linearly fitting score drop vs.\ zoom level reveals a slope close to zero, indicating that we may generally expect a system's MOS to possibly drop by up to around an entire point if we re-evaluate it in a finer-grained context, regardless of the zoom level.  Interestingly, Blizzard Challenge 2011 system G had the largest drop in MOS in both the 6\% test and the 12\% test, becoming the worst-ranked synthesis system in both tests, indicating that perhaps this system's ranking in the original BVCC test results may have been unreliable.

\section{Conclusions}
\label{sec:conclusions}

The design of our set of listening tests that progressively ``zoomed in'' on the high-quality region of a large range of synthesis systems shows a clear presence of range-equalizing bias.  The worst system in the smallest group of top systems dropped by a full 1.28 MOS points, emphasizing the strong contextual effects of MOS-based listening tests.  We can advise that if researchers want to identify more significant differences between systems that are close together in quality, then it is better to keep the overall range of quality of systems in the entire listening test fairly small.  However, it is also important to consider that this may substantially lower the ratings for the worst systems in the group, even if they are overall of relatively high quality.  

One limitation of this study is that when we chose the subsets of the top systems, our choices were based on the original ranking of all 187 systems in BVCC, which we found in Section \ref{sec:correl} may not represent the ``true'' ranking of the systems.  Many of the differences between systems of similar quality were not significant, and ranking order changed upon closer inspection.

We would like to conclude by observing one interesting phenomenon -- the naturalness of Tacotron 2 \cite{taco2} appears to have gotten {\em worse} over the years!  The original paper reported a MOS of 4.53 in 2018 with ``quality close to that of natural human speech,'' whereas a study from 2020 \cite{hayashi2019espnettts} reported a MOS of 4.20, and more recent results from 2022 \cite{styletts} reported a MOS for of 3.01.  Of course, it's not at all valid to make direct comparisons between MOS tests with completely different listeners, test sentences, training data, configurations, vocoders, and even rating scale increments.  However, this general decrease in MOS indicates the possibility of range-equalizing bias, and that an unintentional ``zooming in'' may be taking place as speech synthesis technology improves.  It is therefore of utmost importance for speech synthesis researchers to be mindful of their choices of comparison systems and how they may affect MOS results.  It is also important when interpreting MOS results to be aware of range-equalizing bias, and to understand that what may appear to be a large difference between systems may in fact be a small difference which has been magnified by a narrow range of quality of synthesized samples in the listening test.

In future work, we will conduct additional listening tests such as A/B and MUSHRA in order to better understand how different tests relate to one another.  We will also investigate whether finer-grained listening test data can help MOS predictors to better distinguish between similar systems and to generalize better.  We will also investigate zooming in on the lower- and middle-quality regions of MOS, as well as the effect of different rating scales such as ones using increments of 0.5.

\section{Acknowledgements}

We would like to thank the organizers of the Blizzard Challenges and Voice Conversion Challenges, and the authors of ESPnet-TTS, for making their audio samples freely available.  This study is supported by JST CREST Grant Number JPMJCR18A6 and by MEXT KAKENHI grant 21K11951.

\bibliographystyle{IEEEtran}
\bibliography{mybib}

\end{document}